\documentclass[aps,prd,twocolumn,superscriptaddress,showpacs]{revtex4}

\usepackage{graphicx}
\usepackage{dcolumn}
\usepackage{bm}
\usepackage{rotating}
\usepackage{epsf}
\usepackage{subfigure} 
\usepackage{lineno}
\usepackage{setspace}
\usepackage{amsmath}
\usepackage{bm}
\def\D0bar{\ensuremath{\overline{D}{}^0}}
\def\mum{\ensuremath{\ \mu \mathrm{m}}}

\def\ps{\ensuremath{\ \mathrm{ps}}}

\begin{document}
\title{Measurement of the ${\bm{B_s^0}}$ Lifetime in  Fully and Partially Reconstructed  ${\bm{B_s^0 \to D_s^-(\phi \pi^-) X}}$ Decays in 
${\bm{\overline{p}p}}$ Collisions at ${\bm{\sqrt{s} = 1.96}}$ TeV\\ \vspace*{2.0cm}}
\affiliation{Institute of Physics, Academia Sinica, Taipei, Taiwan 11529, Republic of China} 
\affiliation{Argonne National Laboratory, Argonne, Illinois 60439, USA} 
\affiliation{University of Athens, 157 71 Athens, Greece} 
\affiliation{Institut de Fisica d'Altes Energies, Universitat Autonoma de Barcelona, E-08193, Bellaterra (Barcelona), Spain} 
\affiliation{Baylor University, Waco, Texas 76798, USA} 
\affiliation{Istituto Nazionale di Fisica Nucleare Bologna, $^z$University of Bologna, I-40127 Bologna, Italy} 
\affiliation{University of California, Davis, Davis, California 95616, USA} 
\affiliation{University of California, Los Angeles, Los Angeles, California 90024, USA} 
\affiliation{Instituto de Fisica de Cantabria, CSIC-University of Cantabria, 39005 Santander, Spain} 
\affiliation{Carnegie Mellon University, Pittsburgh, Pennsylvania 15213, USA} 
\affiliation{Enrico Fermi Institute, University of Chicago, Chicago, Illinois 60637, USA}
\affiliation{Comenius University, 842 48 Bratislava, Slovakia; Institute of Experimental Physics, 040 01 Kosice, Slovakia} 
\affiliation{Joint Institute for Nuclear Research, RU-141980 Dubna, Russia} 
\affiliation{Duke University, Durham, North Carolina 27708, USA} 
\affiliation{Fermi National Accelerator Laboratory, Batavia, Illinois 60510, USA} 
\affiliation{University of Florida, Gainesville, Florida 32611, USA} 
\affiliation{Laboratori Nazionali di Frascati, Istituto Nazionale di Fisica Nucleare, I-00044 Frascati, Italy} 
\affiliation{University of Geneva, CH-1211 Geneva 4, Switzerland} 
\affiliation{Glasgow University, Glasgow G12 8QQ, United Kingdom} 
\affiliation{Harvard University, Cambridge, Massachusetts 02138, USA} 
\affiliation{Division of High Energy Physics, Department of Physics, University of Helsinki and Helsinki Institute of Physics, FIN-00014, Helsinki, Finland} 
\affiliation{University of Illinois, Urbana, Illinois 61801, USA} 
\affiliation{The Johns Hopkins University, Baltimore, Maryland 21218, USA} 
\affiliation{Institut f\"{u}r Experimentelle Kernphysik, Karlsruhe Institute of Technology, D-76131 Karlsruhe, Germany} 
\affiliation{Center for High Energy Physics: Kyungpook National University, Daegu 702-701, Korea; Seoul National University, Seoul 151-742, Korea; Sungkyunkwan University, Suwon 440-746, Korea; Korea Institute of Science and Technology Information, Daejeon 305-806, Korea; Chonnam National University, Gwangju 500-757, Korea; Chonbuk National University, Jeonju 561-756, Korea} 
\affiliation{Ernest Orlando Lawrence Berkeley National Laboratory, Berkeley, California 94720, USA} 
\affiliation{University of Liverpool, Liverpool L69 7ZE, United Kingdom} 
\affiliation{University College London, London WC1E 6BT, United Kingdom} 
\affiliation{Centro de Investigaciones Energeticas Medioambientales y Tecnologicas, E-28040 Madrid, Spain} 
\affiliation{Massachusetts Institute of Technology, Cambridge, Massachusetts 02139, USA} 
\affiliation{Institute of Particle Physics: McGill University, Montr\'{e}al, Qu\'{e}bec, Canada H3A~2T8; Simon Fraser University, Burnaby, British Columbia, Canada V5A~1S6; University of Toronto, Toronto, Ontario, Canada M5S~1A7; and TRIUMF, Vancouver, British Columbia, Canada V6T~2A3} 
\affiliation{University of Michigan, Ann Arbor, Michigan 48109, USA} 
\affiliation{Michigan State University, East Lansing, Michigan 48824, USA}
\affiliation{Institution for Theoretical and Experimental Physics, ITEP, Moscow 117259, Russia}
\affiliation{University of New Mexico, Albuquerque, New Mexico 87131, USA} 
\affiliation{Northwestern University, Evanston, Illinois 60208, USA} 
\affiliation{The Ohio State University, Columbus, Ohio 43210, USA} 
\affiliation{Okayama University, Okayama 700-8530, Japan} 
\affiliation{Osaka City University, Osaka 588, Japan} 
\affiliation{University of Oxford, Oxford OX1 3RH, United Kingdom} 
\affiliation{Istituto Nazionale di Fisica Nucleare, Sezione di Padova-Trento, $^{aa}$University of Padova, I-35131 Padova, Italy} 
\affiliation{LPNHE, Universite Pierre et Marie Curie/IN2P3-CNRS, UMR7585, Paris, F-75252 France} 
\affiliation{University of Pennsylvania, Philadelphia, Pennsylvania 19104, USA}
\affiliation{Istituto Nazionale di Fisica Nucleare Pisa, $^{bb}$University of Pisa, $^{cc}$University of Siena and $^{dd}$Scuola Normale Superiore, I-56127 Pisa, Italy} 
\affiliation{University of Pittsburgh, Pittsburgh, Pennsylvania 15260, USA} 
\affiliation{Purdue University, West Lafayette, Indiana 47907, USA} 
\affiliation{University of Rochester, Rochester, New York 14627, USA} 
\affiliation{The Rockefeller University, New York, New York 10065, USA} 
\affiliation{Istituto Nazionale di Fisica Nucleare, Sezione di Roma 1, $^{ee}$Sapienza Universit\`{a} di Roma, I-00185 Roma, Italy} 

\affiliation{Rutgers University, Piscataway, New Jersey 08855, USA} 
\affiliation{Texas A\&M University, College Station, Texas 77843, USA} 
\affiliation{Istituto Nazionale di Fisica Nucleare Trieste/Udine, I-34100 Trieste, $^{ff}$University of Trieste/Udine, I-33100 Udine, Italy} 
\affiliation{University of Tsukuba, Tsukuba, Ibaraki 305, Japan} 
\affiliation{Tufts University, Medford, Massachusetts 02155, USA} 
\affiliation{University of Virginia, Charlottesville, VA  22906, USA}
\affiliation{Waseda University, Tokyo 169, Japan} 
\affiliation{Wayne State University, Detroit, Michigan 48201, USA} 
\affiliation{University of Wisconsin, Madison, Wisconsin 53706, USA} 
\affiliation{Yale University, New Haven, Connecticut 06520, USA} 
\author{T.~Aaltonen}
\affiliation{Division of High Energy Physics, Department of Physics, University of Helsinki and Helsinki Institute of Physics, FIN-00014, Helsinki, Finland}
\author{B.~\'{A}lvarez~Gonz\'{a}lez$^v$}
\affiliation{Instituto de Fisica de Cantabria, CSIC-University of Cantabria, 39005 Santander, Spain}
\author{S.~Amerio}
\affiliation{Istituto Nazionale di Fisica Nucleare, Sezione di Padova-Trento, $^{aa}$University of Padova, I-35131 Padova, Italy} 

\author{D.~Amidei}
\affiliation{University of Michigan, Ann Arbor, Michigan 48109, USA}
\author{A.~Anastassov}
\affiliation{Northwestern University, Evanston, Illinois 60208, USA}
\author{A.~Annovi}
\affiliation{Laboratori Nazionali di Frascati, Istituto Nazionale di Fisica Nucleare, I-00044 Frascati, Italy}
\author{J.~Antos}
\affiliation{Comenius University, 842 48 Bratislava, Slovakia; Institute of Experimental Physics, 040 01 Kosice, Slovakia}
\author{G.~Apollinari}
\affiliation{Fermi National Accelerator Laboratory, Batavia, Illinois 60510, USA}
\author{J.A.~Appel}
\affiliation{Fermi National Accelerator Laboratory, Batavia, Illinois 60510, USA}
\author{A.~Apresyan}
\affiliation{Purdue University, West Lafayette, Indiana 47907, USA}
% ...
\collaboration{CDF Collaboration\footnote{With visitors from $^a$University of Massachusetts Amherst, Amherst, Massachusetts 01003,
$^b$Istituto Nazionale di Fisica Nucleare, Sezione di Cagliari, 09042 Monserrato (Cagliari), Italy,
$^c$University of California Irvine, Irvine, CA  92697, 
$^d$University of California Santa Barbara, Santa Barbara, CA 93106
$^e$University of California Santa Cruz, Santa Cruz, CA  95064,
$^f$CERN,CH-1211 Geneva, Switzerland,
$^g$Cornell University, Ithaca, NY  14853, 
$^h$University of Cyprus, Nicosia CY-1678, Cyprus, 
$^i$University College Dublin, Dublin 4, Ireland,
$^j$University of Fukui, Fukui City, Fukui Prefecture, Japan 910-0017,
$^k$Universidad Iberoamericana, Mexico D.F., Mexico,
$^l$Iowa State University, Ames, IA  50011,
$^m$University of Iowa, Iowa City, IA  52242,
$^n$Kinki University, Higashi-Osaka City, Japan 577-8502,
$^o$Kansas State University, Manhattan, KS 66506,
$^p$University of Manchester, Manchester M13 9PL, England,
$^q$Queen Mary, University of London, London, E1 4NS, England,
$^r$Muons, Inc., Batavia, IL 60510,
$^s$Nagasaki Institute of Applied Science, Nagasaki, Japan, 
$^t$National Research Nuclear University, Moscow, Russia,
$^u$University of Notre Dame, Notre Dame, IN 46556,
$^v$Universidad de Oviedo, E-33007 Oviedo, Spain, 
$^w$Texas Tech University, Lubbock, TX  79609, 
$^x$Universidad Tecnica Federico Santa Maria, 110v Valparaiso, Chile,
$^y$Yarmouk University, Irbid 211-63, Jordan,
$^{gg}$On leave from J.~Stefan Institute, Ljubljana, Slovenia, 
}}
\noaffiliation

\begin{abstract}
We present a measurement of the $B_s^0$ lifetime in fully and partially reconstructed $B_s^0 \to D_s^-(\phi \pi^-) X$ decays in
1.3 fb$^{-1}$ collected in $p\overline{p}$ 
collisions at $\sqrt{s} = 1.96$ TeV by the CDF II 
detector at the Fermilab Tevatron.  We measure
$
    \tau(B_s^0)  =  1.518 \pm 0.041 \ (\mathrm{stat.}) \pm 0.027 \ (\mathrm{syst.})\ps. 
$
The ratio of this result and the world average $B^0$ lifetime yields $\tau(B_s^0)/\tau(B^0)=0.99\pm0.03$, which is in agreement
with recent theoretical predictions. 
\end{abstract}

\maketitle

\par In the spectator model of heavy hadron decay, the lifetimes of all $b$-hadrons are equal, independent of the flavor of
the lighter quarks bound to the $b$ quark.  Using the heavy-quark expansion~\cite{HQE,Bigi} in the calculation of the width, spectator quark 
interactions enter in higher order $(\Lambda_{QCD}/m_b)^3$ terms  where $m_b$ is the mass of the 
$b$ quark and $\Lambda_{QCD}$ is the energy scale of the QCD interactions within the hadron.
This leads to the lifetime
hierarchy 
$\tau(B_s^0)\cong \tau(B^0) < \tau(B^+)$.  
Theoretical results
predict $\tau(B^+)/\tau(B^0)=1.06\pm0.02$ and  $\tau(B_s^0)/\tau(B^0)=1.00\pm0.01$~\cite{theory1,theory2}. 
The world averages for the corresponding experimental numbers are $1.071\pm0.009$ and  
$0.965\pm0.017$, respectively~\cite{PDG}.  
The precision of our knowledge of the $B_s^0$ lifetime is much less than for the $B^0$ and $B^+$ lifetimes, and therefore,
a more precise measurement would be useful, both in general and for comparison with theoretical calculations.
Such a measurement is especially warranted since the agreement on the lifetime ratio between theory and experiment is only fair.

In this Letter, we present a measurement of the $B_s^0$ lifetime in flavor-specific decay modes.  
The data come from $p\overline{p}$ collisions at $\sqrt{s} = 1.96$ TeV
at the Fermilab Tevatron.  This analysis
is based on an integrated luminosity of 1.3 fb$^{-1}$ collected
by the CDF II detector between February 2002 and November 2006.
This sample yields
 more than 1100 fully-reconstructed $B_s^0 \to D_s^-\pi^+$ candidates with $D_s^- \to \phi\pi^-$
and $\phi \to K^+K^-$ after online and offline selection~\cite{conjugation}.
In addition, the sample 
reconstructed as $B_s^0 \to D_s^- \pi^+$ includes partially-reconstructed (PR) $B_s^0$ 
candidates that are used in this lifetime measurement and more than double the number of $B_s^0$ 
candidates available for analysis.
One such PR decay is $B_s^0 \to D_s^- \rho^+$ with $\rho^+ \to \pi^+ \pi^0$ where the $\pi^0$ 
is not reconstructed.   
The inclusion of PR decays introduces an uncertainty in the momentum measurement of a given candidate.
However, a correction to the proper decay time has been estimated, and the total uncertainty on the lifetime measurement
is improved by the use of the PR final states.

\par The CDF II detector is described in detail in Ref.~\cite{CDF}.  The detector elements relevant for this
analysis are the silicon vertex detectors ~\cite{SVXII, L00, ISL} and the central drift chamber (COT)~\cite{COT}.  
The silicon detectors consist of 7 or 8 layers of microstrip silicon sensors covering the pseudo-rapidity~\cite{coordinates}
 range $| \eta | <2.0$. 
The COT is an open cell drift chamber covering $| \eta | <1.0$.  
Both the COT and silicon vertex detectors are immersed in a uniform 1.4 T axial magnetic field with the field axis parallel to the proton beam.

\par A data sample enriched in hadronic $B$ decays is selected with a three-level trigger system that searches for 
tracks displaced from the primary vertex.  At level 1, patterns of hits in the COT are 
identified as tracks by the extremely fast tracker (XFT)~\cite{XFT}.
At level 2, the silicon vertex trigger~\cite{SVT} associates a set of silicon hits with the 
XFT tracks and improves track measurement precision.
The trigger requires each event to contain a pair of charged particle tracks, 
each having transverse momentum $p_T \geq 2$ GeV/$c$ and transverse
impact parameter $d_0$ in the range $d_0 \in$  [120 $\mu$m, 1 mm],
where $d_0$ is defined as the distance of closest approach between the 
particle trajectory and the beamline, measured in the transverse plane.  
The opening angle between the tracks' trajectories 
($\Delta\phi$ in the plane transverse to the beam) must be between
$2^\circ$ and $90^\circ$, and their intersection must be at least 200 $\mu$m from
the interaction point, as measured in the plane transverse to the beam direction.
At level 3 track reconstruction is performed entirely in software, with the full precision of the tracking system available, and
the level 1 and 2 requirements are confirmed.  
These trigger requirements preferentially select events containing long-lived particles and sculpt the proper time distribution
of the particles that are accepted for analysis.
CDF also employs two more restrictive triggers that require the tracks in the trigger pair to have opposite charges,
individual $p_T \geq 2(2.5)$ GeV/$c$, and the scalar sum $p_T \geq 5.5(6.5)$ GeV/$c$.

\par We reconstruct $B_s^0\to D_s^-\pi^+$ candidates  
(where $B_s^0$ and $D_s^-$ imply $B_s^0$ candidates and $D_s^-$ candidates) 
by first identifying  $D_s^-\to\phi(K^-K^+)\pi^-$  from tracks with $p_T>350$ MeV/$c$ 
using the invariant mass requirements  
$\left|m(K^-K^+) - 1020.5\right| < 7.5$ MeV/$c^2$ and $\left|m(K^-K^+\pi^-) - 1968.3\right| < 20$ MeV/$c^2$.
The $D_s^-$ daughter tracks must satisfy a three-dimensional vertex fit.  We then combine each $D_s^-$
with a positively charged track with $p_T > 1.0$ GeV/$c$ to form a $B_s^0\to D_s^-\pi^+$  candidate and require the pair to satisfy an 
additional three-dimensional vertex fit.  We do not constrain the mass of the $\phi$ or $D_s^-$ in this fit.
The decay length of the $B_s^0$ is measured with respect to the event's primary vertex and must satisfy requirements
on the following quantities:  
the decay length of the $B_s^0$ projected along the transverse momentum, $L_{xy}(B_s^0)>450\mum$, and its significance,
$L_{xy}(B_s^0)/\sigma_{L_{xy}}(B_s^0)>5$; 
the transverse distance between the $B_s^0$ and $D_s^-$ decay points is greater than 0; 
the transverse impact parameter of the $B_s^0$, $\left|d_0(B_s^0)\right|<60\mum$; and the 
significance of the longitudinal impact parameter, $\left|z_0(B_s^0)/\sigma_{z_0}(B_s^0)\right|<3$. 
Both fits for the  $B_s^0$ and $D_s^-$ vertices must have reasonable goodness-of-fit values when considering
only the track parameters measured in the transverse plane.

\par To further separate $B_s^0$ mesons from backgrounds with similar topologies, we require
the transverse momentum of the $B_s^0$, $p_T(B_s^0)>5.5$ GeV/$c$, and
the angular separation between the $D_s^-$ and the $\pi$ from the $B_s^0$, $\Delta R(D_s^-,\pi_B) = \sqrt{(\Delta \eta)^2 +(\Delta \phi)^2}<1.5$.
We require the isolation of the $B_s^0$ to be greater than 0.5, defined as
$p_T(B_s^0)$ divided by  the scalar sum of the transverse momenta of all the tracks in a cone of 
$\Delta R<1$ around the $B_s^0$.  
We tighten the requirement on the mass of the $D_s^-$ ($\left|m - 1968.3\right| < 12$ MeV/$c^2$) and veto
$D_s^-$ candidates consistent with $D^{*-} \rightarrow \D0bar \pi^-$ where the $\D0bar$ decays to $K^+\pi^-\pi^0$.
The $D^*$ veto is accomplished by taking the $D_s^-$ daughter tracks 
($K^-K^+\pi^-$), assigning the pion mass to the negative kaon, and requiring $\Delta m = m(K^+\pi^-\pi^-) - m(K^+\pi^-) > 180$ MeV/$c^2$.
We also require that the decay contain two reconstructed tracks satisfying the level 2 trigger requirements.

\par The simulated data samples used in this analysis consist of single $b$-hadrons generated by 
{\sc{bgenerator}}~\cite{bgen1,bgen2}
with $p_T$ spectra
consistent with NLO QCD and decayed with {\sc{evtgen}}~\cite{evtgen}.
Full detector and trigger simulations are performed.
 The simulated $B$ candidates are reconstructed with the same procedure and the same
selection as the data candidates.  We reweight the simulated sample to match the data distributions for 
$p_T(B)$ and trigger scenario mixture.

\par The lifetime of the $B_s^0$ meson is determined from two sequential fits.  
The first is a fit to the invariant mass distribution of 
candidates reconstructed as 
$D_s^-\pi^+$ used to determine the fractions of the total number of events found in the various decay modes.
These fractions are then used as fixed inputs in the second fit to 
the proper decay time distributions of the $B_s^0$.  
The uncertainties on the fractions returned by the mass fit are treated as sources
of systematic uncertainty.

\par The mass fit is an unbinned maximum likelihood fit to the invariant mass of the candidate reconstructed as 
$D_s^-\pi^+$ with $m_{B}^{rec} \in$ [4.85,6.45] GeV/$c^2$.  The mass fit components can 
be characterized as coming from one of three possible sources: single $b$-hadrons, real-$D_s^-$+track background,
and fake-$D_s^-$+track background. The mass probability distribution functions (PDFs)
 for single $b$-hadrons were obtained from simulation, with an additional small 
shift and resolution smearing to bring the simulated $B_s^0$-peak central value and width into agreement with data. 
The single-$b$ modes were separated in the fit as follows:
$    B_{s}^0 \to D_{s}^-\pi^+(n\gamma)  $,
$               B_{s}^0 \to D_{s}^{\pm}K^{\mp}  $,
$            B_{s}^0 \to D_{s}^-\rho^+  $,
$         B_{s}^0 \to D_{s}^{-*}\pi^+  $,
$         B_{s}^0 \to D_{s}^{-(*)}X  $,
$         B^{0}/B^{-} \to D^{-}X  $,
$       B^{0} \to D_{s}^{-(*)}\pi^+ + D_{s}^{-(*)}K^+  $, and 
$   \Lambda_{b} \to \Lambda_{c}X  $.  
The $D_{s}^{\pm}K^{\mp}/D_{s}^-\pi^+$ ratio was constrained to the results of Ref.~\cite{dkdpi}.

\begin{figure}[!htb]
  \begin{center}
    \includegraphics[width=3.2in]{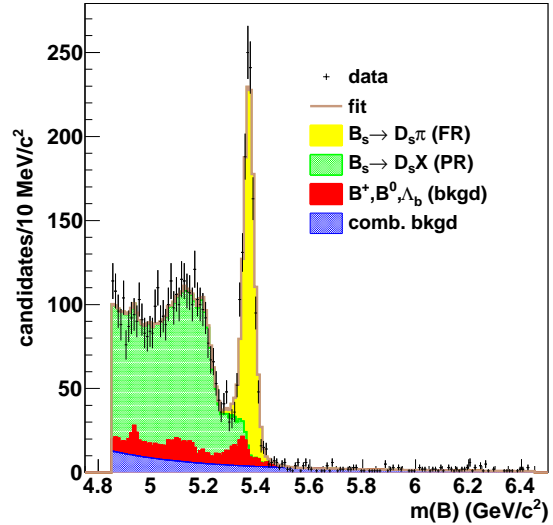}
  \end{center}
  \caption{Mass distribution for candidates reconstructed as  $B_s^0 \to D_s^-\pi^+$ with fit projections overlaid.}
  \label{fig:massfitBsDsPhipi}
\end{figure}

Real-$D_s^-$+track backgrounds consist
of a real $D_s^-$, produced promptly or from
a $b$-hadron decay, plus an additional track produced in the event.
The mass PDF for these events is obtained from an auxiliary fit to the wrong-sign sample, which consists of  
data events reconstructed as $D_s^- \pi^-$ and sideband subtracted in the $D_s^-$ mass.  
The mass PDF for the fake-$D_s^-$+track background  is obtained from an auxiliary fit to the $D_s^-$ sidebands. 
The results of the mass fit are shown in Fig.~\ref{fig:massfitBsDsPhipi} with various modes combined for plotting only. 
The real-$D_s^-$+track background
and fake-$D_s^-$+track background components are drawn together as the ``combinatorial background''.

For the lifetime fit, the
variable of interest is the proper decay time, defined as $ct \equiv
( L_{xy}(B_s^0) \cdot m_{B}^{rec})/p_T(B_s^0)$.  The
reconstructed mass $m_{B}^{rec}$ is used instead of the world average $B_s^0$ mass.  A
salient feature of this analysis is the treatment of partially-reconstructed $B_s^0$ mesons as signal events that contribute to the
lifetime measurement.  Since in the partially-reconstructed cases
$L_{xy}(B_s^0)$, $m_{B}^{rec}$, and ${p_T(B_s^0)}$ are extracted from candidates
that are missing particles after reconstruction or have the wrong mass
assignment for a daughter particle, a multiplicative correction factor $K$ to
the decay time is needed.  $K$ is defined as  
$K = \left[ p_T(B_s^0) \cdot m_B^{true}\right] / \left[ p_T^{true}(B_s^0) \cdot m_B^{rec} \cdot \cos \theta_{\mathrm{PR}}\right]$
where $ \theta_{PR}$ is the angle in the $x-y$ plane between the true momentum of the $B_s^0$ and the
momentum of the partially-reconstructed $B_s^0$. 
Because the ratio   $m_{B}^{rec}/{p_T(B_s^0)}$ is numerically very close to the ratio $m_B^{true} /p_T^{true}(B_s^0)$,
this choice 
of $ct$ definition forces the $K$ factor distributions to be centered near $K=1$ with widths of a few percent.  
The $K$ factor distributions are determined with simulation.

\par The lifetime of the $B_s^0$ meson is determined from an unbinned likelihood fit to the $B_s^0$ candidates 
with invariant masses in the range [5.00, 5.45] GeV/$c^2$.  
There are three main types of lifetime fit components that will be described in the following paragraphs: 
fully-reconstructed $B_s^0$, partially-reconstructed $B_s^0$, and
backgrounds.
The treatment of each component depends on its decay structure and whether it can provide information about the
$B_s^0$ lifetime. 

\par Fully-reconstructed (FR) modes where all of the $B_s^0$ daughter particles are included with the correct mass assignment in the construction
of the $B_s^0$ candidate are the first type of lifetime fit component.   
The only FR mode in this analysis is the $D_s^-\pi^+$.
The core functional form of the FR PDF is 
an exponential with decay constant $c\tau(B_s^0)$ convoluted with a Gaussian
resolution function with width $\sigma$:
\begin{equation}
P_{\mathrm{FR}}(ct)= \left[ \frac{1}{c\tau} e^{\textstyle\frac{-ct'}{c\tau}}
 \otimes_{t'} \frac{1}{\sqrt{2\pi}\sigma}e^{\textstyle\frac{-(ct-ct')^2}{2\sigma^2}}
   \right]
   \cdot \mbox{eff}(ct)
   \label{frPDF}
\end{equation}
A multiplicative ``efficiency curve'' accounts for the trigger 
and analysis selection criteria:

\[
\mathrm{eff}(ct) = 
\sum_{i=1}^{3} N_i \cdot (ct-\beta_i)^2 \cdot e^\frac{-ct}{\tau_i} \cdot \theta(\beta_i)
\]
The shape  parameters ($\sigma$, $\beta_i$, $N_i$ and $\tau_i$) of the PDF are determined in
a fit to a simulated $B_s^0$ sample where the lifetime used for generation is known.  
All the parameters for the PDF are then fixed and only $\tau(B_s^0)$ is varied in the final fit to the data.  
As we depend on the simulation of the displaced-track trigger, we use a data sample of $J/\psi \to \mu^+\mu^-$ 
decays collected with a di-muon trigger to
assess the accuracy of this assumption and assign a ``trigger simulation'' systematic uncertainty based on these studies.
The partially-reconstructed, {\sc{photos}}-modeled
$D_s^-\pi(n\gamma)$ 
decays~\cite{photos} are combined with the FR $D_s^-\pi$, as the momentum carried by the photon is small and their lifetime distribution
is extremely close to the FR one.  
This simplification is considered as a possible source
of systematic uncertainty. 

\par Partially-reconstructed  modes either neglect $B_s^0$ daughter particles in the construction
of the $B_s^0$ candidate or assign them an incorrect mass. 
$B_s^0 \rightarrow D_s^-K^+$, $D_s^-\rho^+$, $D_s^{*-}\pi^+$, and other decay modes partially reconstructed under the  $D_s^- \pi^+$ hypothesis can 
also contribute to the $B_s^0$ lifetime measurement. 
The PR PDF  is similar to the FR PDF of Eq.~(\ref{frPDF}) 
with an additional convolution with the $K$ factor distribution for each
mode.  There are separate efficiency curve parameters for each mode, again determined from fits to simulated events.

\par The backgrounds in the lifetime fit can either come from decays of $b$-hadrons other than the $B_s^0$
({\it e.g.}, $B^0/B^-\rightarrow D^-X$, $B^0\rightarrow D_s^- X$, and $\Lambda_b \rightarrow \Lambda_c X$),
or they can come from real-$D_s^-$+track and fake-$D_s^-$+track combinations. 
The PDFs for the former modes are derived from fits to simulated $B^0$, $B^-$ and $\Lambda_b$ samples, while
the ones for the latter combinatorial backgrounds come from the two proxies available:
the $B_s^0$ upper sideband taken from the $m_B^{rec}$ interval [5.7, 6.4] GeV/$c^2$ and the $D_s^-$ sidebands.
The $D_s^-$ sidebands are taken from the  $m_B^{rec}$ interval [5.0, 5.45] GeV/$c^2$ 
and the $m_{D}^{rec}$ interval [1.924,1.939]$\bigcup$[1.999,2.014] GeV/$c^2$. 
Both proxies contain a mixture of fake $D_s^-+$track events and real $D_s^-+$track events, where a real $D_s^-$ can
be poorly reconstructed. 
All the background shape parameters are fixed in the final lifetime fit.

\par The analysis procedure was tested extensively on three control samples: 
$B^0\rightarrow D^- \pi^+$ with $D^- \to K^+\pi^-\pi^-$, 
$B^0\rightarrow D^{*-} \pi^+$ with $D^{*-} \to \D0bar \pi^-$ and $\D0bar \to K^+\pi^-$, and
$B^+\rightarrow\D0bar \pi^+$ with $\D0bar \to K^+\pi^-$  
 before performing the $B_s^0$ fits.   Furthermore, the lifetime fit of the $B_s^0$ was performed using a {\it blind}
approach, i.e. by determining the statistical and systematic uncertainties without knowledge of the fit result itself. 
Good agreement 
with the world average values~\cite{PDG} of the $B^0$ and $B^+$ lifetimes was found.

\par
\begin{figure}[!htb]
   \begin{center}
     \includegraphics[trim = 0mm 2.55in 0mm 0mm, clip, width=3.2in]{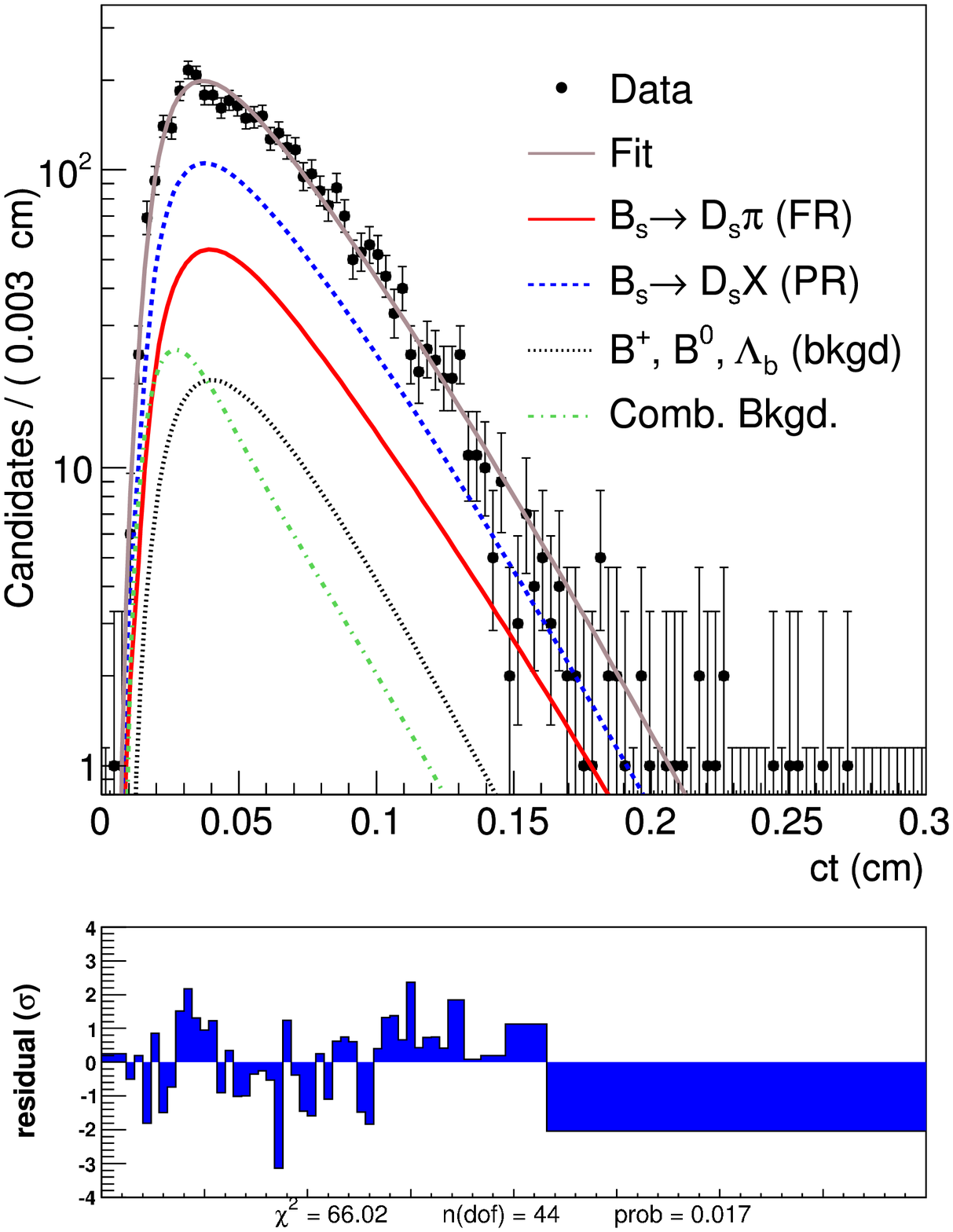}      
     \caption[ct projections of lifetime fit results]{Distribution of $ct$ for candidates reconstructed as $B_s^0 \to D_s^-(\phi \pi^-)\pi^+$ with fit projection overlaid.}  
     \label{fig:BsDsPhipi_ct_results}
   \end{center}
 \end{figure}

 The lifetime of 
$\tau(B_s^0) =  1.518 \pm 0.041 \ (\mathrm{stat.})\ps$ 
is obtained from the full fit.
The fit results are plotted in Fig.~\ref{fig:BsDsPhipi_ct_results}.
The results of the fits performed separately in the FR mass region 
(1.456 $\pm$ 0.067\ps) %
and PR mass region (1.544 $\pm$  0.051\ps) %
agree with each other at a level of 1.0 $\sigma$.

\begin{table}[!htb]
\caption{Summary of sources of systematic uncertainty for the $B_s^0\to D_s^- (\phi \pi^-)X$ lifetime fit. 
The total uncertainty is calculated assuming the individual contributions are uncorrelated.} 
\label{tab:sys}
\begin{center}
\begin{tabular}{lc}\hline \hline
Description                                       & Value (ps)        \\ \hline
Background modeling and fractions                 &  0.019              \\ %
Fixed single-$b$ background $ct$ 		  &  0.003              \\ %
Reweighting for $p_T$ and trigger		  &  0.012              \\ %
Lifetime contribution of $D\pi$ radiative tail	  &  0.002              \\ %
Efficiency curve parameterization		  &  0.002              \\ %
Trigger simulation             			  &  0.014              \\ %
Impact parameter correlation			  &  0.003              \\ %
Detector alignment 			          &  0.003              \\ \hline
Total systematic uncertainty			  &  0.027              \\ \hline\hline
\end{tabular}
\end{center}
\end{table}
\par   We use a Monte Carlo technique to assess the systematic uncertainties.
For each source of systematic uncertainty, we generate 1000 simulated experiments with the number of events in each experiment 
Poisson-distributed around the number of events in data.
The simulated experiments are generated with a non-standard lifetime fit configuration (where the PDFs or numbers of events in the various modes 
are modified to account for the systematic effect) and fit with the default configuration.
The mean biases returned from the fits to the simulated experiments
($\tau_{\mathrm{ret}}-\tau_{\mathrm{gen}}$) are used
to set the size of the systematic uncertainties.  
We consider several sources of systematic
uncertainty:  combinatorial background fraction, modeling of backgrounds
from single $b$-hadron decays, effect of reweighting the full simulations to match the data, 
modeling of the trigger bias as a function of $ct$, 
offline-online impact parameter correlation, accuracy of the trigger simulation in Monte Carlo, and detector alignment.
Table~\ref{tab:sys} contains the final list
of systematic uncertainties for this measurement.  
The largest contribution comes from the uncertainty on the total amount of 
combinatorial background and the amount of promptly produced real-$D_s^-$ background. 

\par The displaced-track trigger, in addition to modifying the accepted decay length distribution from
a simple exponential to the form in Eq.~(\ref{frPDF}), alters the expected mixture of mass eigenstates
$B_{s,L}^0$ and $B_{s,H}^0$ in the flavor-specific $B_s^0 \to D_s^- \pi^+$ decay by
preferentially selecting the longer lived $B_{s,H}^0$.  The size of the imbalance can be calculated 
using the parameters of the efficiency curve and the world 
average of $1/\Gamma = 1.47$ ps~\cite{PDG}.  Our result can be corrected back to a flavor-specific 
lifetime measurement with $\delta \tau(B_s^0) = -0.11 \left(\Delta\Gamma/\Gamma\right)^2$ ps.
Given the world average  $\Delta\Gamma/\Gamma = 0.092^{+0.051}_{-0.054}$~\cite{PDG}, the correction 
would be smaller than our
statistical and systematic uncertainties.  Therefore, we do not correct the central value or assess an additional systematic
uncertainty.

\par  In summary we have measured the $B_s^0$ lifetime using both fully reconstructed $B_s^0 \to D_s^-(\phi \pi^-)\pi^+$
and partially reconstructed $B_s^0 \to D_s^-(\phi \pi^-) X$ decay modes, in a sample with 
1.3 fb$^{-1}$ of integrated luminosity.
We measure 
$ 
 \tau(B_s^0)  =  1.518 \pm 0.041 \ (\mathrm{stat.}) \pm 0.027 \ (\mathrm{syst.})\ps, 
$
which is consistent with theoretical expectations~\cite{theory1, theory2}.

We thank the Fermilab staff and the technical staffs of the participating institutions for their vital contributions. 
This work was supported by the U.S. Department of Energy and National Science Foundation; the Italian Istituto Nazionale di Fisica Nucleare; 
the Ministry of Education, Culture, Sports, Science and Technology of Japan; the Natural Sciences and Engineering Research Council of Canada; 
the National Science Council of the Republic of China; the Swiss National Science Foundation; the A.P. Sloan Foundation; the Bundesministerium 
f\"ur Bildung und Forschung, Germany; the Korean World Class University Program, the National Research Foundation of Korea; the Science and 
Technology Facilities Council and the Royal Society, UK; the Institut National de Physique Nucleaire et Physique des Particules/CNRS; the 
Russian Foundation for Basic Research; the Ministerio de Ciencia e Innovaci\'{o}n, and Programa Consolider-Ingenio 2010, Spain; the Slovak 
R\&D Agency; the Academy of Finland; and the Australian Research Council (ARC).


\begin{thebibliography}{99}
\bibitem{conjugation}
      Charge conjugation is implied throughout this Letter.
\bibitem{coordinates}
      CDF II uses a right-handed coordinate system with the origin at the center of the detector, in which the $z$ axis
      is along the proton direction, the $y$ axis points up, $\theta$ and $\phi$ are the polar and azimuthal angles, 
      and $r$ is the radial distance in the $x–y$ plane. The pseudorapidity $\eta$ is deﬁned as −log tan($\theta$/2).
\end{thebibliography}
\end{document}